\documentclass[a4paper,11pt,final]{article}

\usepackage{amsmath,amsfonts,amssymb,bm,mathptmx,graphicx}
\usepackage[small,bf]{caption}

\DeclareFontFamily{OT1}{pzc}{}
\DeclareFontShape{OT1}{pzc}{m}{it}{<-> s*[1.125] pzcmi7t}{}
\DeclareMathAlphabet{\mathcal}{OT1}{pzc}{m}{it}

\usepackage{hyperref}
\hypersetup{pdftitle={The inactive-active phase transition in the noisy additive (exclusive-or) probabilistic cellular automaton}, pdfauthor={J. Ricardo G. Mendonça}, unicode, extension=pdf, pdfnewwindow, pdftoolbar, pdfmenubar, pdffitwindow=false, pdfstartview={FitH}, colorlinks, linkcolor=blue, citecolor=blue, urlcolor=blue}
\makeatletter
\renewcommand\@makefntext[1]{\leftskip=0.0em\hskip-0.5em\@makefnmark#1}
\makeatother
\def\leq{\leqslant}
\def\geq{\geqslant}
\def\giv{\,|\,}

\def\1{\mbox{\bf 1}}

\begin{document}

\pagestyle{empty}

\renewcommand{\thefootnote}{\fnsymbol{footnote}}

\begin{titlepage}

\begin{center}

{\Large {\bf The inactive-active phase transition in the noisy additive \\ (exclusive-or) probabilistic cellular automaton}}

\vspace{6ex}

{\large {\bf J. Ricardo G. Mendon\c{c}a}\footnote{Email: {\tt \href{mailto:jricardo@ime.usp.br}{\nolinkurl{jricardo@usp.br}}}}}

\vspace{1ex}

{\it \mbox{Complex Systems Group, Escola de Artes, Ci\^{e}ncias e Humanidades, Universidade de S\~{a}o Paulo} \\ \mbox{Rua Arlindo Bettio 1000, Ermelino Matarazzo, 03828-000 S\~{a}o Paulo, SP, Brazil}}

\vspace{3ex}

{\large \bf Abstract \\}

\vspace{2ex}

\parbox{130mm}
{We investigate the inactive-active phase transition in an array of additive (exclusive-or) cellular automata under noise. The model is closely related with the Domany-Kinzel probabilistic cellular automaton, for which there are rigorous as well as numerical estimates on the transition probabilities. Here we characterize the critical behavior of the noisy additive cellular automaton by mean field analysis and finite-size scaling and show that its phase transition belongs to the directed percolation universality class of critical behavior. As a by-product of our analysis, we argue that the critical behavior of the noisy elementary CA~90 and 102 (in Wolfram's enumeration scheme) must be the same. We also perform an empirical investigation of the mean field equations to assess their quality and find that away from the critical point (but not necessarily very far away) the mean field approximations provide a reasonably good description of the dynamics of the PCA.

\vspace{2ex}

{\noindent}{\bf Keywords}: {Additive cellular automata} $\cdot$ {diluted CA 90} $\cdot$ {Domany-Kinzel cellular automaton} $\cdot$ {universality class} $\cdot$ {phase transition} $\cdot$ {mean field approximation}

\vspace{2ex}

{\noindent} {\bf PACS 2010}:
02.50.Ga $\cdot$ 05.70.Fh $\cdot$ 64.60.De $\cdot$ 64.60.Ht

\vspace{2ex}

{\noindent}{\bf Journal ref}.:
\href{http://dx.doi.org/10.1142/S0129183116500169}
{{\it Int. J. Mod. Phys. C\/} {\bf 27}\,(2), 1650016 (2016)}}

\end{center}

\end{titlepage}

\pagestyle{plain}


\section{\label{intro}Introduction}

Cellular automata (CA) are discrete-space, discrete-time synchronous (states change simultaneously everywhere) deterministic dynamical systems that map symbols from a finite alphabet into the same alphabet according to local (finite range) rules. The origins of CA date back at least to the early days of the modern computer era ($\sim$1940--1950), when they were conceived as model systems for simple self-reproducing, self-repairing organisms and, by extension, logical elements operating in parallel and memory storage devices, among others \cite{jvn,wolfram,toff,sarkar,kari}

Probabilistic cellular automata (PCA), in turn, are CA that evolve by local rules that depend on the realization of some random variable \cite{reliable,vaser,toom90,toom95} In computer science, PCA are used, e.g., to model the impacts of faulty elements (like reading heads, logical circuitry, or neurons) on the computational capability of CA. Besides serving as model systems for the analysis of computation---both applied and theoretical, digital or biological---under noise, PCA have also been playing a significant role in the elucidation of some deep issues in equilibrium and nonequilibrium statistical mechanics \cite{domany,kinzel,grinstein,pierre,lebowitz} We also mention that, currently, CA and PCA are very useful tools in the modeling of biological and ecological complex systems \cite{edels,levin,silva,landman}

In this work we investigate the very simple noisy additive (exclusive-or) PCA with respect to its inactive-active phase transition. The rigorous determination of the parameters for which the noisy additive PCA is ergodic is an open problem \cite{mair} In the jargon of statistical mechanics, this question amounts to determining whether the noisy additive PCA displays a phase transition between a phase with a single stationary state devoid of active cells (the inactive phase) and a phase where the stationary state has a finite density of active cells (the active phase) and, if yes, to estimate the critical parameters of the model. We partially answer this question by showing that the noisy additive PCA corresponds to the Domany-Kinzel PCA \cite{domany,kinzel} in one of its parameter subspaces. We have organized this paper as follows. In Sec.~\ref{model}, the noisy additive and the Domany-Kinzel PCA are described, and we show how they are equivalent to each other and to other known CA and PCA in the literature. In Sec.~\ref{meanfield} we provide simple analytical approximations to the critical point of the noisy additive PCA by standard mean field techniques and in Sec.~\ref{mcarlo} we supplement the analysis with Monte Carlo simulations and finite-size scaling analysis, obtaining relatively precise estimates for its critical point and critical exponents. In Sec.~\ref{assess} we make an attempt to assess how good mean field approximations are for the noisy additive PCA by measuring (numerically) the discrepancies between some exact quantities and the same quantities in the mean field approximation. Finally, in Sec.~\ref{summary} we summarize our results and suggest a few directions for further investigation.


\section{\label{model}The noisy additive cellular automata}

\subsection{\label{pxor}The $p$-XOR PCA}

The noisy additive cellular automaton is a two-state PCA with state space given by $\Omega_{\Lambda} = \{0,1\}^{\Lambda}$, with $\Lambda \subseteq \mathbb{Z}$ a finite array of $|\Lambda|=L \geq 1$ cells under periodic boundary conditions $i+L \equiv i$, and transition function $\Phi: \Omega_{\Lambda} \to \Omega_{\Lambda}$ that given the state $\bm{x}^{t} = (x_{1}^{t}, x_{2}^{t}, \ldots, x_{L}^{t})$ of the PCA at instant $t \in \mathbb{N}$ determines the state $x_{i}^{t+1} = [\Phi(\bm{x}^{t})]_{i} = \phi(x^{t}_{i}, x^{t}_{i+1})$ of the PCA at instant $t+1$ according to the following rule: with probability $p \in [0,1]$, $x_{i}^{t+1} = x_{i}^{t}~\oplus~x_{i+1}^{t}$ (the exclusive-or operation), otherwise $x_{i}^{t+1}=0$. In terms of conditional probabilities $W(x_{i} \giv x_{i}, x_{i+1})$, we have the set of transition rules $W(1 \giv 0,0) = 0$, $W(1 \giv 0,1) = W(1 \giv 1,0) = p$, and $W(1 \giv 1,1) = 0$, with $W(0 \giv \,\cdot\,, \,\cdot\,) = 1-W(1 \giv \,\cdot\,, \,\cdot\,)$. This model was first considered by Vasershtein back in 1969 \cite{vaser}, among other models that were later to be reintroduced in the literature. We have also considered this PCA before, where the phase transition and critical parameters of a closely related asynchronous interacting particle system were established by a transfer matrix-like technique \cite{non}. We dub the noisy additive PCA the $p$-XOR PCA. Its rule table over a symmetric neighborhood of radius 1 appears in table~\ref{tab:pxor}.

The $p$-XOR PCA interpolates between the trivial CA $\phi(x^{t}_{i}, x^{t}_{i+1})=0$ at $p=0$ and the elementary CA~102 (in Wolfram's enumeration scheme \cite{wolfram}) at $p=1$, being a mixture of the two (a dilution'' of CA~102) for $0 < p < 1$. Such probabilistic mixtures of deterministic CA have been considered before \cite{hans,bhatta,p182q200}; in the notation of \cite{p182q200}, the $p$-XOR PCA becomes the $p102$--$(1-p)0$ PCA. If $p$ is small, we can reasonably expect that the $p$-XOR PCA will eventually converge to the absorbing state $\bm{0}=(0,0,\ldots,0)$ devoid of active cells, the corresponding invariant measure being denoted by $\bm{\delta}_{0}$. It is not known, rigorously, whether there exists a critical value $p^{*}$ such that for $p > p^{*}$ the invariant measures of the $p$-XOR PCA become translation-invariant convex combinations of the form $\alpha\bm{\delta}_{0} + (1-\alpha)\bm{\mu}_{p}$, with $0 < \alpha < 1$ and $\bm{\mu}_{p}$ the measure that puts mass on configurations with density $0 < \mu_{p}(1) < 1$.

We will show that the $p$-XOR PCA is closely related with the Domany-Kinzel (DK) PCA \cite{domany,kinzel}, which displays several inactive-active-type phase transitions and for which there are rigorous as well as numerical estimates on the transition probabilities. In the next section we describe the DK PCA and how the two PCA are related to each other and to other known CA and PCA in the literature.

\begin{table}[t]
\centering
\caption{Rule table for the noisy additive PCA. The first row lists the initial neighborhood and the other two rows give the probability at which the central cell reaches the state given in the leftmost column.}
{\begin{tabular}{ccccccccc}
\hline\hline
  {}  & 111 & 110 & 101 & 100 & 011 & 010 & 001 & 000 \\
\noalign{\smallskip}\hline\noalign{\smallskip}
  0:  & 1 & $1-p$ & $1-p$ & 1 & 1 & $1-p$ & $1-p$ & 1 \\
  1:  & 0 & $p$ & $p$ & 0 & 0 & $p$ & $p$ & 0 \\
\hline\hline
\end{tabular}}
\label{tab:pxor}
\end{table}

\subsection{\label{dkpca}The Domany-Kinzel PCA}

The Domany-Kinzel (DK) PCA is a two- (sometimes \mbox{three-}) parameter PCA originally introduced to investigate the relationship between $d$-dimensional nonequilibrium models and $d+1$-dimensional equilibrium models, casting some light on this difficult subject \cite{domany,kinzel,grinstein,pierre,lebowitz,mair}. Its rule table appears in table~\ref{tab:dk}. The two-dimensional parameter space of the DK PCA encompasses several well-known, archetypal models in discrete mathematics and theoretical physics, among them the directed site ($p_{1}=p_{2}$) and the directed bond ($p_{2}=2p_{1}-p_{1}^{2}$) percolation processes, the exactly solvable compact directed percolation process, which is equivalent to the zero temperature Glauber-Ising model, the voter model, and diffusing-annihilating random walks ($p_{1}=1/2$, $p_{2}=1$), and a couple of other models, like the elementary CA~90 ($p_{1}=1$, $p_{2}=0$)\footnote{Refs. \cite{haye,stav} incorrectly state that the line $p_{2}=0$ of the DK PCA corresponds to the diluted elementary CA~18.} and Stavskaya's PCA ($p_{1} = p_{2} = 1-\varepsilon$, the noise parameter of Stavskaya's PCA) \cite{domany,kinzel,grinstein,pierre,lebowitz,mair,liggett,marro,haye,stav}.

\begin{table}[t]
\centering
\caption{Rule table for the Domany-Kinzel PCA. Same notation as in table~\ref{tab:pxor}.}
{\begin{tabular}{ccccccccc}
\hline\hline
{} &  111  &  110  &  101  &  100  &  011  & 010 &  001  & 000 \\
\noalign{\smallskip}\hline\noalign{\smallskip}
0: & $1-p_2$ & $1-p_1$ & $1-p_2$ & $1-p_1$ & $1-p_1$ & 1 & $1-p_1$ & 1 \\
1: & $p_2$ & $p_1$ & $p_2$ & $p_1$ & $p_1$ & 0 & $p_1$ & 0 \\
\hline\hline
\end{tabular}}
\label{tab:dk}
\end{table}

From tables~\ref{tab:pxor} and~\ref{tab:dk}, we see that the $p$-XOR and the DK PCA do not immediately relate to each other. Rule tables~\ref{tab:pxor} and \ref{tab:dk} are compatible only at the point $p=p_{1}=p_{2}=0$, that corresponds to the trivial CA~0. Indeed, in the notation of \cite{p182q200}, we have that the $p$-XOR PCA is the $p102$--$(1-p)0$ PCA, while the DK$(p_{1},p_{2})$ PCA becomes, at special points, PCA DK$(p,0)=$ $p90$--$(1-p)0$, DK$(p,p)=$ $p250$--$(1-p)0$, and DK$(0,p)=$ $p160$--$(1-p)0$, and each of these PCA does not relate with PCA $p102$--$(1-p)0$ except at the trivial point $p=0$.

\begin{figure}[t]
\centering
\includegraphics[viewport = 240 120 600 480, scale=0.50, clip=true]{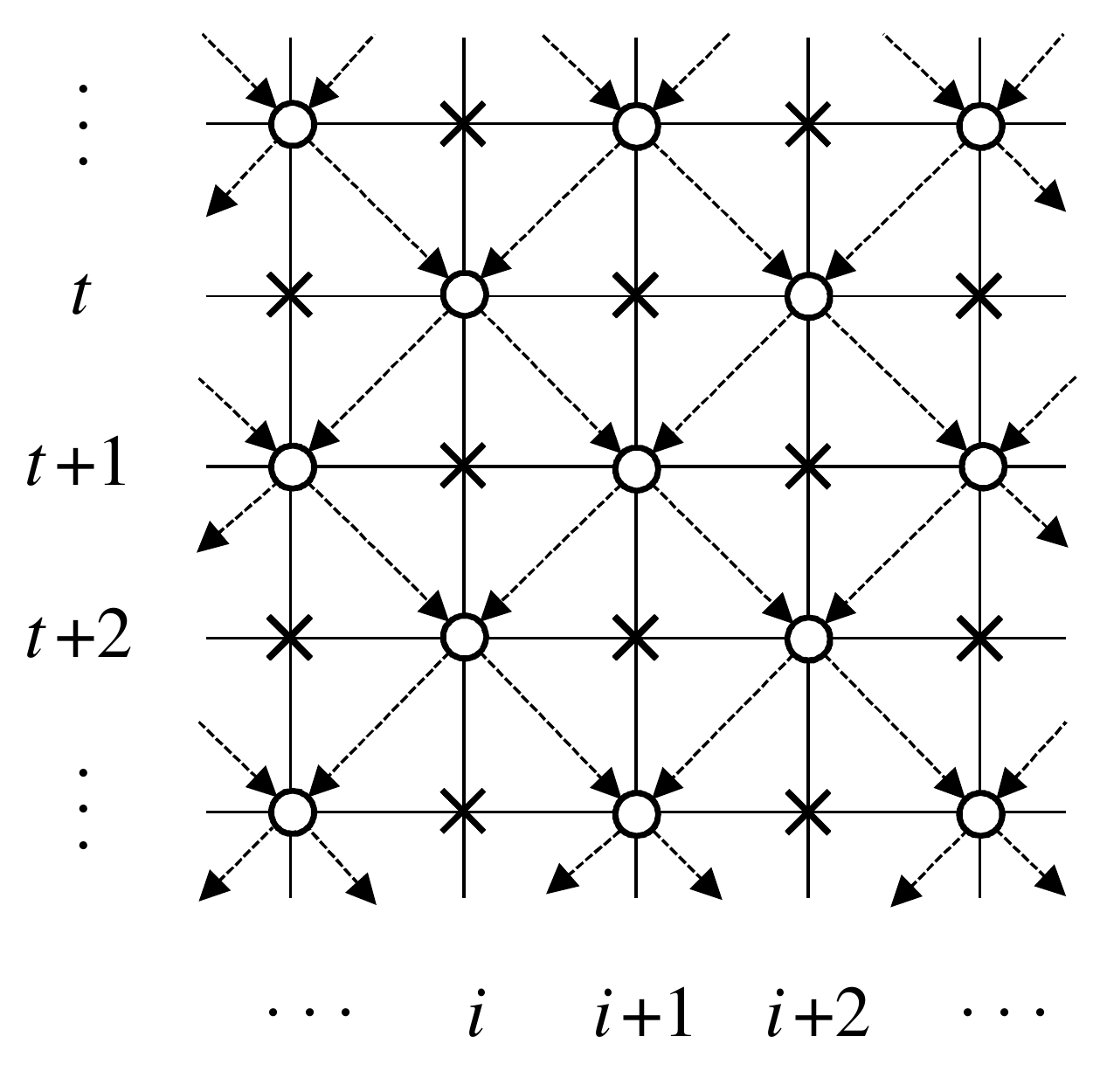}
\caption{\label{fig:sub}The dynamics of the DK PCA occurs in two separated sublattices that do not interact. In the figure, the circles together with the dashed arrows display one of the sublattices.}
\end{figure}

Note, however, that the dynamics of the DK PCA occurs in two separated sublattices that do not interact, each of which evolves by a rule that can be written as $x_{i}^{t+1}=$ $\phi(x_{i}^{t}, x_{i+1}^{t})$, see figure~\ref{fig:sub}. If we recast the DK PCA in one of its sublattices as a PCA over a symmetric neighborhood of radius 1 again, we obtain the rule table given in table~\ref{tab:dksub}. Now, comparing the DK PCA dynamics in one of its sublattices with the $p$-XOR PCA dynamics, we see that the $p$-xor PCA corresponds to the DK PCA on the line $p_{2}=0$. On this line, it is known that the ``diluted'' CA~90 and the DK PCA coincide and that they display an inactive-active phase transition at $p_{1}^{*} \simeq 0.81$ \cite{domany,kinzel,martins,penna,bagnoli,weitz,kemper}.

We could have simplified our discussion above by looking at the rules of the $p$-XOR and the sublattice DK PCA in their asymmetric versions $x_{i}^{t+1} = \phi(x_{i}^{t}, x_{i+1}^{t})$ and noticing that they are equal when $p_{2}=0$. But the discussion using symmetric neighborhoods of radius 1 indicates that the critical behavior of the diluted CA~90 (or PCA $p90$--$(1-p)0$), represented by the DK PCA at the point $p_{2}=0$, and the diluted CA~102 (or PCA $p102$--$(1-p)0$), represented by the $p$-XOR PCA, must be the same. The fact that these two CA/PCA may display the same critical behavior seems to have gone unnoticed in the literature so far. Note, however, that it has been found that some properties may differ whether one uses the full lattice or the sublattice version of the DK PCA \cite{atman}. It should be remarked, however, that for the cases investigated in \cite{atman} there is an accompanying change in the neighborhood that changes the nature of the rules (the full lattice case includes the cell itself in the rule); this can be clearly seen if we blow the resulting rules up to a symmetric neighborhood of radius 1, as we did for the $p$-XOR PCA---we would then discover that the two models investigated there are compatible only on the trivial point $p_{1}=p_{2}=0$.

\begin{table}[t]
\centering
\caption{Rule table for the Domany-Kinzel PCA in one of its sublattices recast as a PCA over a symmetric neighborhood of radius 1. Same notation as in tables~\ref{tab:pxor}--\ref{tab:dk}.}
{\begin{tabular}{ccccccccc}
\hline\hline
{} &  111  &  110  &  101  & 100 &  011  &  010  &  001  & 000 \\
\noalign{\smallskip}\hline\noalign{\smallskip}
0: & $1-p_2$ & $1-p_1$ & $1-p_1$ & 1 & $1-p_2$ & $1-p_1$ & $1-p_1$ & 1 \\
1: & $p_2$ & $p_1$ & $p_1$ & 0 & $p_2$ & $p_1$  & $p_1$ & 0 \\
\hline\hline
\end{tabular}}
\label{tab:dksub}
\end{table}


\section{\label{meanfield}Mean field analysis}

The Markovian dynamics of the probability distribution $P_{t}(\bm{x})$ of the states $\bm{x} \in \Omega_{\Lambda}$ of the PCA is governed by the equation
\begin{equation}
\label{ptx}
P_{t+1}(\bm{x}') = \sum_{\bm{x}}W(\bm{x}' \giv \bm{x})P_{t}(\bm{x}),
\end{equation}
where the summation runs over all $\bm{x} \in \Omega_{\Lambda}$ and $W(\bm{x}' \giv \bm{x}) \geq 0$ is the conditional probability for the transition $\bm{x}$ $\to$ $\bm{x}'$ to occur in one time step. When the cells of the PCA are updated simultaneously and independently we have
\begin{equation}
\label{www}
W(\bm{x}' \giv \bm{x}) =
\prod_{i=1}^{L}W_{i}(x_{i}' \giv \bm{x}), \quad {\rm with} \quad
\sum_{x_{i}'}W_{i}(x_{i}' \giv \bm{x}) = 1.
\end{equation}

For the $p$-XOR PCA, $W_{i}({x_{i}'} \giv \bm{x})=$ $W({x_{i}'} \giv x_{i}, x_{i+1})$, independent of $i$---the PCA is homogeneous in space. The marginal probability distribution $P_{t}(x_{1}, \ldots, x_{n})$ of observing a block of $n$ consecutive cells in state $(x_{1}, \ldots, x_{n})$ is obtained from $P_{t}(\bm{x})$ by summing it over the variables $x_{n+1}$, \ldots, $x_{L}$. From (\ref{ptx}) and (\ref{www}), we derive that the dynamics of $P_{t}(x_{1}, \ldots, x_{n})$ is given by
\begin{equation}
\label{margs}
P_{t+1}(x_{1}', \ldots, x_{n}') = \sum_{x_{1},\, \cdots,\, x_{n+1}}
\bigg[ \prod_{j=i}^{n} W(x_{j}' \giv x_{j}, x_{j+1}) \bigg] 
P_{t}(x_{1}, \ldots, x_{n+1}).
\end{equation}
We see that the probability of observing $n$ consecutive cells in a given state at instant $t+1$ depends on the probabilities of observing the state of $n+1$ cells at instant $t$.

To proceed with the calculations, either we solve the full set of $L$ coupled equations (\ref{margs}) exactly or truncate the hierarchy of equations (\ref{margs}) at some point to get a closed set of equations. The simplest approximation is obtained by taking
\begin{equation}
\label{approx}
P_{t}(x_{1}, x_{2}) \approx P_{t}(x_{1})P_{t}(x_{2}),
\end{equation}
while higher order approximations can be obtained by the generalized approximation
\begin{equation}
\label{mfield}
P_{t}(x_{1}, \ldots, x_{n}) \approx
\frac{P_{t}(x_{1}, \ldots, x_{n-1})\, P_{t}(x_{2}, \ldots, x_{n})}
{P_{t}(x_{2}, \ldots, x_{n-1})}.
\end{equation}
This approximation has been described in the context of cellular automata in \cite{gutowitz}. For pragmatic expositions of the technique see \cite{benav,tania,emilio,rgdk,hons,mjo,rechtman}, while further developments and applications can be found in \cite{hons,rechtman,maps,fuks}.

We note the following features of the $p$-XOR PCA and equations (\ref{margs}) and (\ref{mfield}):
\begin{itemize}
\item[($i$)]Since $W(\,\cdot\, \giv x_{1}, x_{2}) = W(\,\cdot\, \giv x_{2}, x_{1})$, the PCA is reflection-symmetric and the marginal probability distributions observe $P(x_{1}, x_{2}, \ldots, x_{n}) = P(x_{n}, x_{n-1}, \ldots, x_{1})$;
\item[($ii$)]The cluster approximation (\ref{mfield}) is consistent with the reflection-symmetry of the marginal probability distributions, i.\,e., the approximations of $P(x_{1}, x_{2}, \ldots, x_{n})$ and $P(x_{n}, x_{n-1}, \ldots, x_{1})$ according to (\ref{mfield}) coincide;
\item[($iii$)]If we take $P(1,1,\ldots,1)$ on the left-hand side of (\ref{margs}), only the terms $P(1,0,1,\ldots)$ and $P(0,1,0,\ldots)$ appear on the right-hand side, because both $W(1 \giv 0,0)$ and $W(1 \giv 1,1)$ are zero. Moreover, for even number or arguments $P(1,0,\ldots,1,0)=$ $P(0,1,\ldots,0,1)$ by property~($i$) above, further simplifying the mean field equations.
\end{itemize}
We now proceed with the mean field approximation to the full probability distribution of configuration of the $p$-XOR PCA up to the level of two cells.

\paragraph{\label{single}Single-cell approximation.}

From table~\ref{tab:pxor} and equations (\ref{margs})--(\ref{mfield}), the single-cell approximation for $P_{t+1}(x=1)$ reads
\begin{equation}
\label{onexact}
\begin{split}
P_{t+1}(1) &= pP_{t}(0,1) + pP_{t}(1,0) = 2pP_{t}(0,1) \\
      &\approx 2pP_{t}(0)P_{t}(1) = 2p[1-P_{t}(1)]P_{t}(1).
\end{split}
\end{equation}
In the stationary state, $P_{t+1}(1)=P_{t}(1)$ and we obtain
\begin{equation}
\label{onecell}
P(1) = 2p[1-P(1)]P(1),
\end{equation}
with solutions $P(1)=0$ and $P(1)=(2p-1)/2p$. The first solution corresponds to the single state $\bm{0}=(0,\ldots,0)$ devoid of active cells. The other solution is negative (unphysical) for $p<1/2$ and positive for $p>1/2$, thus predicting a phase transition at $p^{*}=1/2$ from the inactive state to a state with density $0 \leq P(1)=(2p-1)/2p \leq 1/2$ of active cells. Note that in this approximation, $P(1)=1/2$ at $p=1$, which coincides with the exact value of the stationary density of active cells of CA~90 \cite{wolfram}, which is the DK PCA at $p_{1}=1$, $p_{2}=0$. The approximation $P(x_{1}, x_{2}, x_{3}) \approx P(x_{1})P(x_{2})P(x_{3})$ of \cite[(3.4)]{wolfram} also estimates the stationary value $P(1)=1/2$ for CA~102 (recall that the $p$-XOR PCA is, in the notation of \cite{p182q200}, the mixed PCA $p102$--$(1-p)0$).

\paragraph{\label{pair}Two-cell approximation.}

Since $P_{t}(0,1)+P_{t}(1,1)=$ $P_{t}(1)$ and $P_{t}(1,0)=$ $P_{t}(0,1)$, we only need to write equations for $P_{t}(1)$ and $P_{t}(1,1)$ to characterize the two-cell mean field approximation of the $p$-XOR PCA. The equation for $P_{t}(1)$ is (\ref{onexact}), and from table~\ref{tab:pxor} and equation (\ref{margs}) we obtain
\begin{equation}
\label{twoexact}
P_{t+1}(1,1) = p^{2}P_{t}(0,1,0) + p^{2}P_{t}(1,0,1). 
\end{equation}
The approximation (\ref{mfield}) provides
\begin{equation}
\label{twoapp}
\begin{split}
P_{t}(0,1,0) &\approx \frac{P_{t}(0,1)P_{t}(1,0)}{P_{t}(1)}, \\
P_{t}(1,0,1) &\approx \frac{P_{t}(1,0)P_{t}(0,1)}{P_{t}(0)},
\end{split}
\end{equation}
such that, in the two-cell approximation,
\begin{equation}
\label{twocell}
P_{t+1}(1,1) \approx p^{2}\frac{[P_{t}(1)-P_{t}(1,1)]^{2}}{P_{t}(1)[1-P_{t}(1)]}
\end{equation}
We want to solve for $P(1)$, the order parameter'' of the model. From the exact part of equation (\ref{onexact}) we must have $P(1,1) = (2p-1)P(1)/2p$ in the stationary state. Imposing the stationarity condition $P_{t+1}(1,1) = P_{t}(1,1)$ in (\ref{twocell}) and plugging in the expression for $P(1,1)$ in terms of $P(1)$ furnishes the solution
\begin{equation}
\label{twostat}
P(1) = \frac{3p-2}{4p-2}
\end{equation}
for the density of active cells in the stationary state of the $p$-XOR PCA in the two-cell mean field approximation. This result predicts an inactive-active phase transition at the critical point $p^{*}=2/3$; again, we have $P(1)=1/2$ at $p=1$.

The two-cell mean field approximations for the dynamics of the $p$-XOR PCA provide the lower-bound $p^{*} \geq 2/3$ for the critical value of the model. Upper bounds are more difficult to obtain \cite{liggett}. Although it is possible to improve the lower bound by pushing the mean field approximation to higher orders, the equations soon become somewhat ungainly. We shall proceed to Monte Carlo simulations and finite-size scaling analysis to gain a better understanding of the phase transition of the model.

\begin{figure}[ht]
\centering
\includegraphics[viewport = 280 80 480 520, scale=0.35]{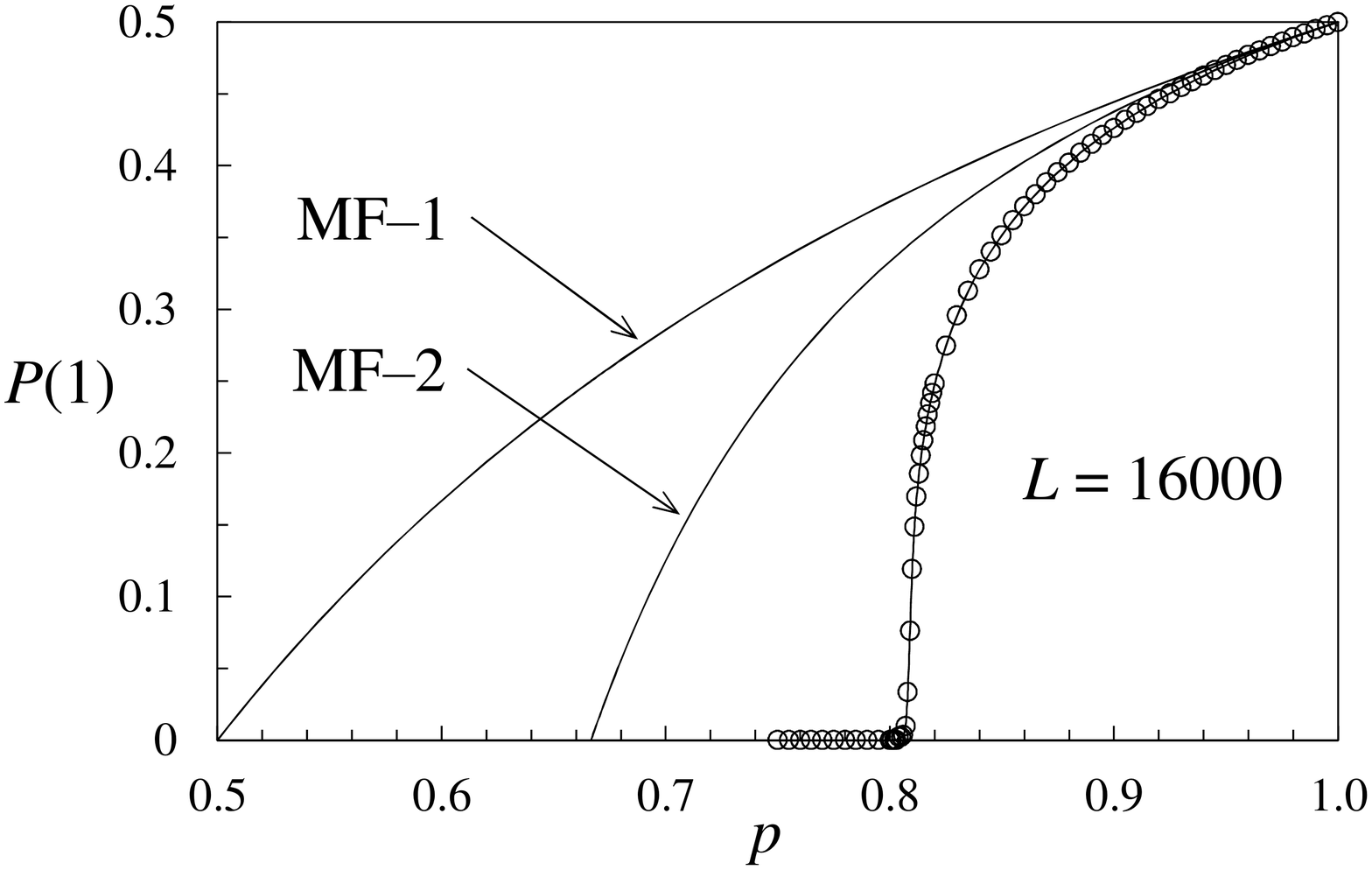}
\caption{\label{fig:rho}Stationary density $P(1)$ of active cells in the one- and two-cell mean-field approximations together with Monte Carlo simulation data for a lattice of $L=16000$ cells initialized randomly with density ${\sim}1/3$ (cf.~Sec.~\ref{mcarlo}). Each symbol in the Monte Carlo curve is an average over $2500$ samples. These data estimate the critical point at $p^{*} \simeq 0.81$.}
\end{figure}


\section{\label{mcarlo}Finite-size scaling analysis}

Our Monte Carlo simulations of the $p$-XOR PCA ran as follows. For a given $p$, the PCA is initialized with each cell $x_{i}^{0}=1$, $1 \leq i \leq L$, drawn independently with probability $1/3$, i.e., $\bm{x}^{0} \sim {\rm Binomial}(L,1/3)$. Stationary state quantities, e.g. the density of active cells $\rho_{L} \equiv P(1) = L^{-1}\sum_{i}x_{i}$, are then sampled after the system is relaxed through $5L$ Monte Carlo steps (MCS), with one MCS equivalent to a synchronous update of the states of all $L$ cells of the automaton. This amount of relaxation proved sufficient to reach the stationary state away from the critical point. Note, however, that except for the data in Fig.~\ref{fig:rho}, our results were obtained from time-dependent simulations, not from averages in the stationary state.

\begin{figure}[ht]
\centering
\begin{tabular}{c}
\includegraphics[viewport = 60 60 750 540, scale=0.32, clip]{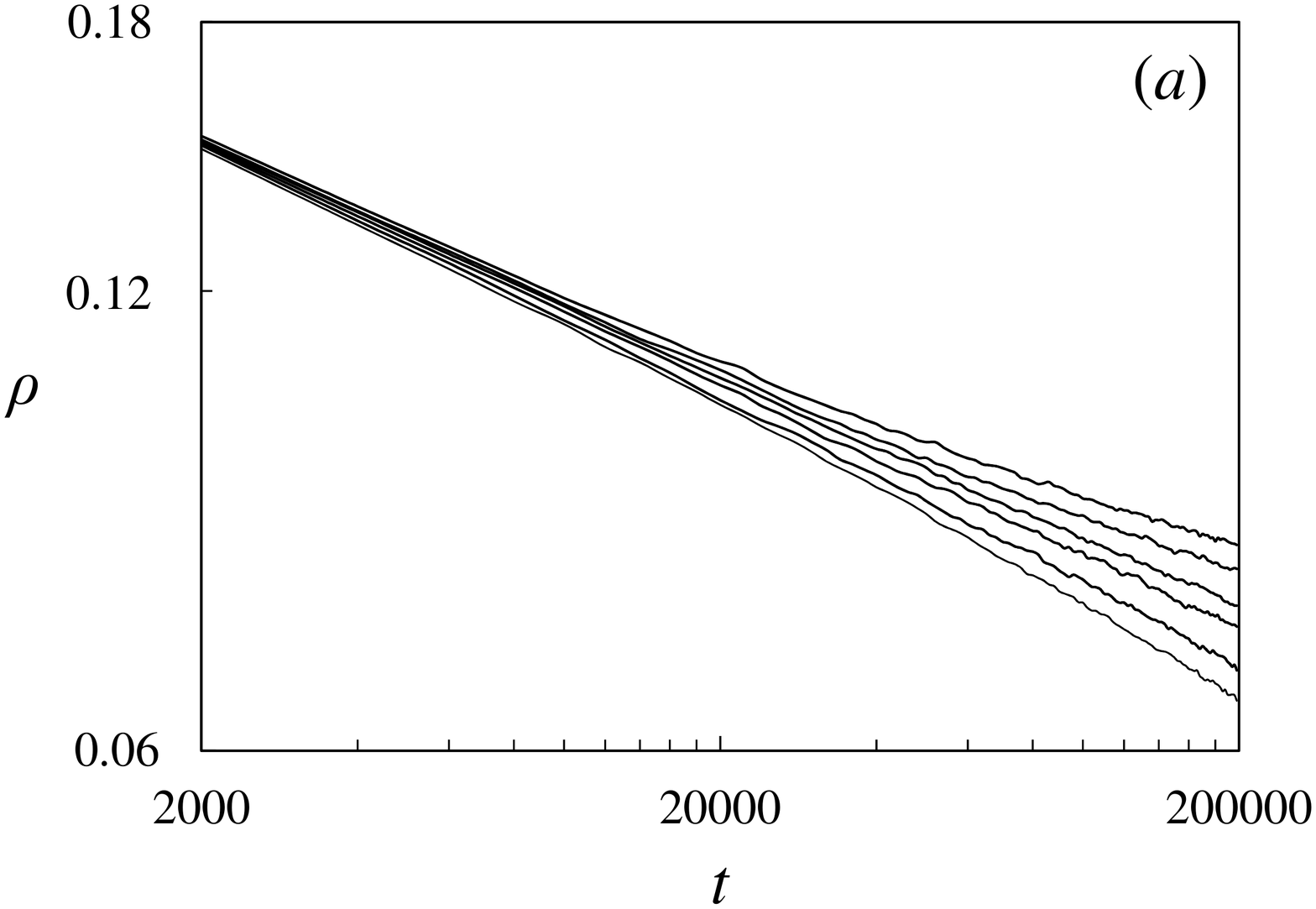} \\
\includegraphics[viewport = 60 60 750 540, scale=0.32, clip]{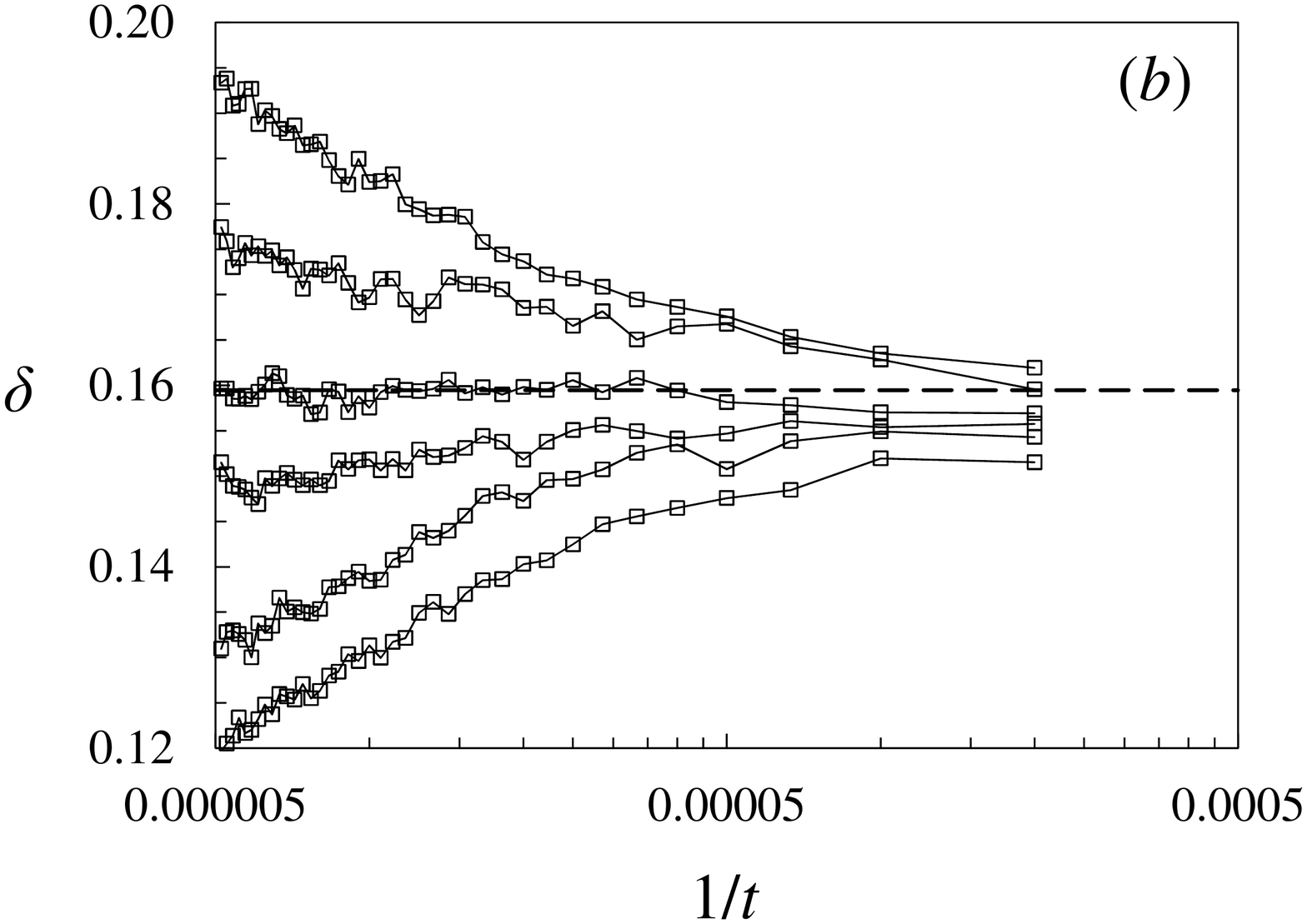}
\end{tabular}
\caption{\label{fig:delta}($a$)~Logarithmic plot of $\rho_{L}(t)$ with $L=16000$ cells. ($b$)~Instantaneous values of $\delta$ obtained from the curves in ($a$). In panel~($a$) we have, from the lowermost curve upwards, $p=0.80930$, $0.80934$, $0.80938$, $0.80942$, $0.80946$, and $0.80950$, while in panel~($b$) this order is reversed. The dashed line in panel~($b$) indicates the best value available for $\delta_{\rm DP}=0.159\,464(6)$.}
\end{figure}

According to the general theory of critical phenomena for equilibrium and nonequilibrium systems \cite{marro,haye,amit}, we can assume that close to the critical point $p_{L}^{*}$ the density $\rho_{L}$ of active cells obeys the finite-size scaling relation
\begin{equation}
\label{scaling}
\rho_{L}(t;\Delta_{L}) \sim t^{-\beta/\nu_{\|}}\,\Phi(\Delta_{L}t^{1/\nu_{\|}},\, t^{\nu_{\perp}/\nu_{\|}}/L),
\end{equation}
with $\Delta_{L} = p-p_{L}^{*} \geq 0$ and $L$ the size of the PCA array. For a very large system, $\rho_{L}(t;\Delta_{L}) \sim$ $t^{-\beta/\nu_{\|}}\,\Phi(\Delta_{L}t^{1/\nu_{\|}})$, with $\Phi(x \ll 1)~\sim$ const and $\Phi(x \gg 1) \sim x^{\beta}$. The investigation of the time-dependent profiles $\rho_{L}(t;\Delta_{L})$ then allows the simultaneous determination of $p_{L}^{*}$ and $\delta = \beta/\nu_{\|}$. Perusal of (\ref{scaling}) and derived relations furnish the other exponents, see, e.g., \cite{haye}.

\begin{figure}[ht]
\centering
\includegraphics[viewport = 60 60 780 540, scale=0.32, clip]{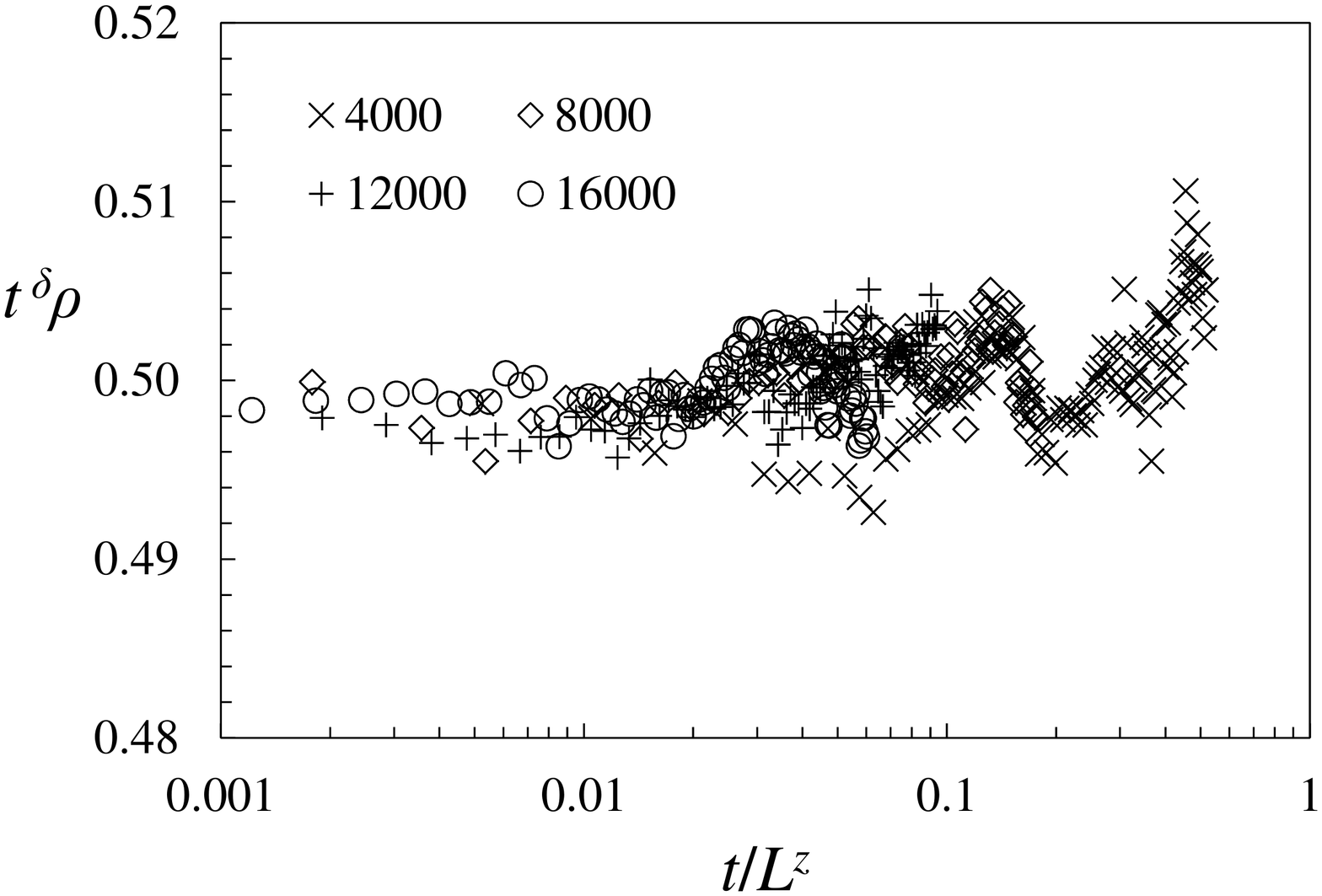}
\caption{\label{fig:ze}Data collapse of the scaled time-dependent density profiles on the critical point $p^{*}=0.80939$ for some $4000 \leq L \leq 16000$. Best collapse was obtained with $\delta=0.158$ and $z=1.55$.}
\end{figure}

Figure~\ref{fig:rho} displays the density profile $\rho_{L}$ for an automaton of $L=16000$ cells in the stationary state. The steep transition about $p_{L}^{*} \sim 0.81$ anticipates a small value for the exponent $\beta$. Note that this rough estimate of $p^{*}$ agrees well with the critical point $0.80 \lesssim p_{1}^{*} \lesssim 0.81$ found before for the DK PCA phase transition along the line $p_{2}=0$ \cite{domany,kinzel,martins,penna,weitz,kemper,rechtman}.

\begin{figure}[t]
\centering
\includegraphics[viewport = 60 60 780 540, scale=0.32, clip]{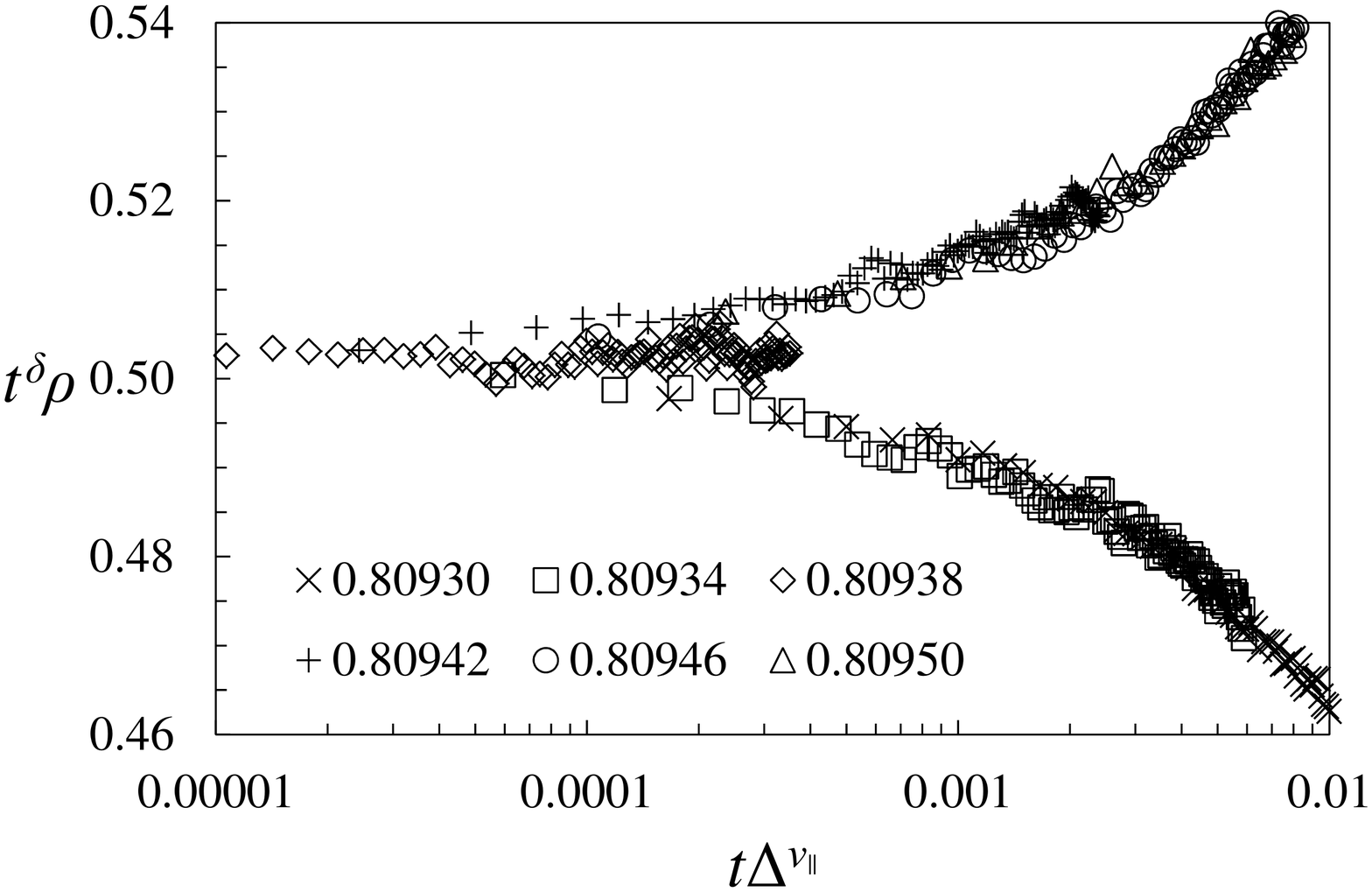}
\caption{\label{fig:nu}Data collapse of the scaled time-dependent density profiles for several different values of $p$. The lower (upper) branches correspond to $p < p^{*}$ ($p > p^{*}$). The best data collapse was obtained with $p^{*}=0.80939$, $\delta=0.159$, and $\nu_{\|}=1.75$.}
\end{figure}

To estimate $p^{*}$ more precisely, we plot $\rho_{L}(t)$ close to $p \simeq 0.81$ for some large $L$. On the critical point, $\rho_{L}(t) \sim t^{-\delta}$ and we can estimate $\delta$ by plotting $\log_{b}[\rho_{L}(t/b)/\rho_{L}(t)]$ against $1/t$ for some small $b$. Our data for $L=16000$ and $b=10$ appear in figure~\ref{fig:delta}, where each curve is an average over 2500 realizations of the process. From these data we extract the estimates $p^{*}=0.80938(4)$ and $\delta=0.159(2)$, where the numbers in parentheses indicate the uncertainty in the last digit of the data. The value for $\delta$ was obtained from extrapolations of the curve at $p=0.80938$, which give $\delta(t)=$ $-12.746\,t^{-1}+0.1593$ with a correlation coefficient $R^{2}=0.1506$, or, alternatively, $\delta(t)=$ $0.1593\,\exp(-80.5\,t^{-1})$ with a correlation coefficient $R^{2}=0.1516$.

The exponent $\nu_{\|}$ can be obtained by plotting $t^{\delta}\rho_{L}(t)$ against $t\Delta_{L}^{\nu_{\|}}$ and tuning $\nu_{\|}$ to achieve data collapse with different $\Delta_{L}$. Figure~\ref{fig:nu} shows the collapsed curves obtained with $p^{*}=0.80939$, $\delta=0.159$, and $\nu_{\|}=1.75$. We could not discern the value of $\nu_{\|}$ more precisely than by $\pm 0.05$. Otherwise, the data collapse is very sensitive to $p^{*}$ and we used this fact to further bracket the critical point to $p^{*}=0.80939(3)$. Combining $\delta=0.159(2)$ and $\nu_{\|}=1.75(5)$ furnishes $\beta=\delta\nu_{\|}=0.278(9)$.

The third independent exponent can be obtained by plotting $t^{\delta}\rho_{L}(t)$ versus $t/L^{z}$ for different $L$ and tuning $z$ until data collapse for some $z$. Since $z = \nu_{\|}/\nu_{\perp}$ by definition, this procedure also gives $\nu_{\perp}$, once $\nu_{\|}$ is known. The finite-size curves appear in figure~\ref{fig:ze}. We found best data collapse with $z=1.55(5)$ and $\delta=0.158$.

Exponents $\delta$, $\nu_{\|}$, and $z$ suffice to determine the universality class of critical behavior of the model, the other exponents following from hyperscaling relations \cite{marro,haye}. The best values available for the critical exponents of the $(1+1)$-dimensional directed percolation process on the square lattice are $\delta_{\rm DP}=0.159\,464(6)$, $\nu_{\|{\rm DP}}=$ $1.733\,847(6)$, $\beta_{\rm DP}=$ $0.276\,486(8)$, and 
$z_{\rm DP}=$ $1.580\,745(10)$ \cite{jensen,precise}. Thus, within the error bars, our estimates of the exponents of the $p$-XOR PCA---namely, $\delta=0.159(2)$, $\nu_{\|}=1.75(5)$, $\beta=\delta\nu_{\|}=0.278(9)$, and $z=1.55(5)$---put its phase transition in the $(1+1)$-dimensional directed percolation universality class of critical behavior.


\section{\label{assess}An assessment of the mean field approximation}

While in the one hand it is hard to obtain rigorous bounds on the distance between the full exact invariant measure for the PCA and the approximations provided by the mean field equations, on the other hand one can worry that the one- and two-cell mean field approximations are not very good away from $p=1$, as it is clear from figure~\ref{fig:rho}. Moreover, although we can reasonably expect that higher order mean field approximations become better and better, it is our experience that the convergence of the approximations to the limit values close to the critical regions is slow with the order of the approximation, making simple extrapolation strategies not very useful---the fact the stationary measure becomes singular (in the limit $L \to \infty$) in the vicinity of a phase transition only makes things tougher.

\begin{figure}[ht!]
\centering
{\begin{tabular}{c}
\includegraphics[viewport = 60 60 780 540, scale=0.32, clip]{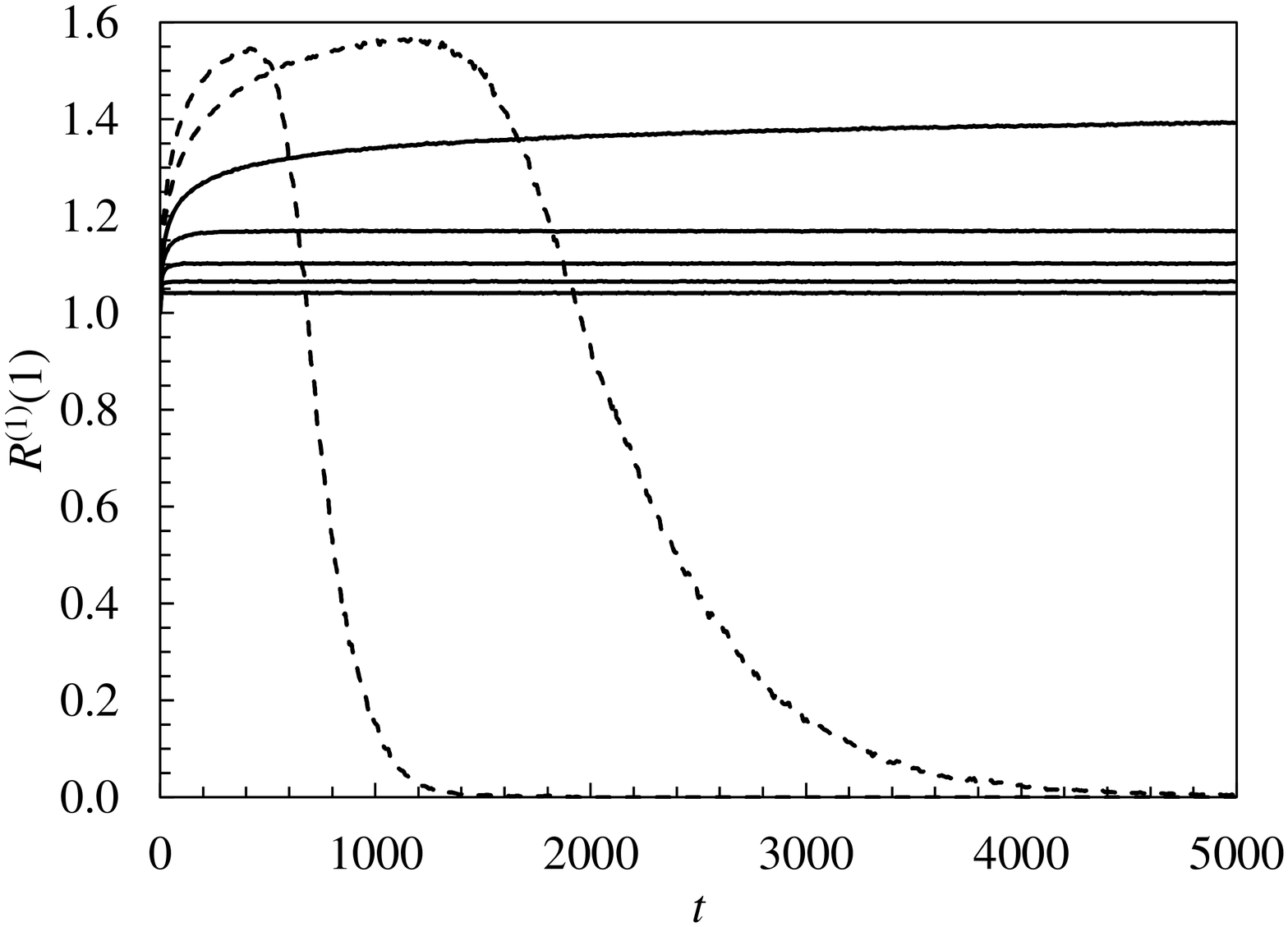}  \\
\includegraphics[viewport = 60 60 780 540, scale=0.32, clip]{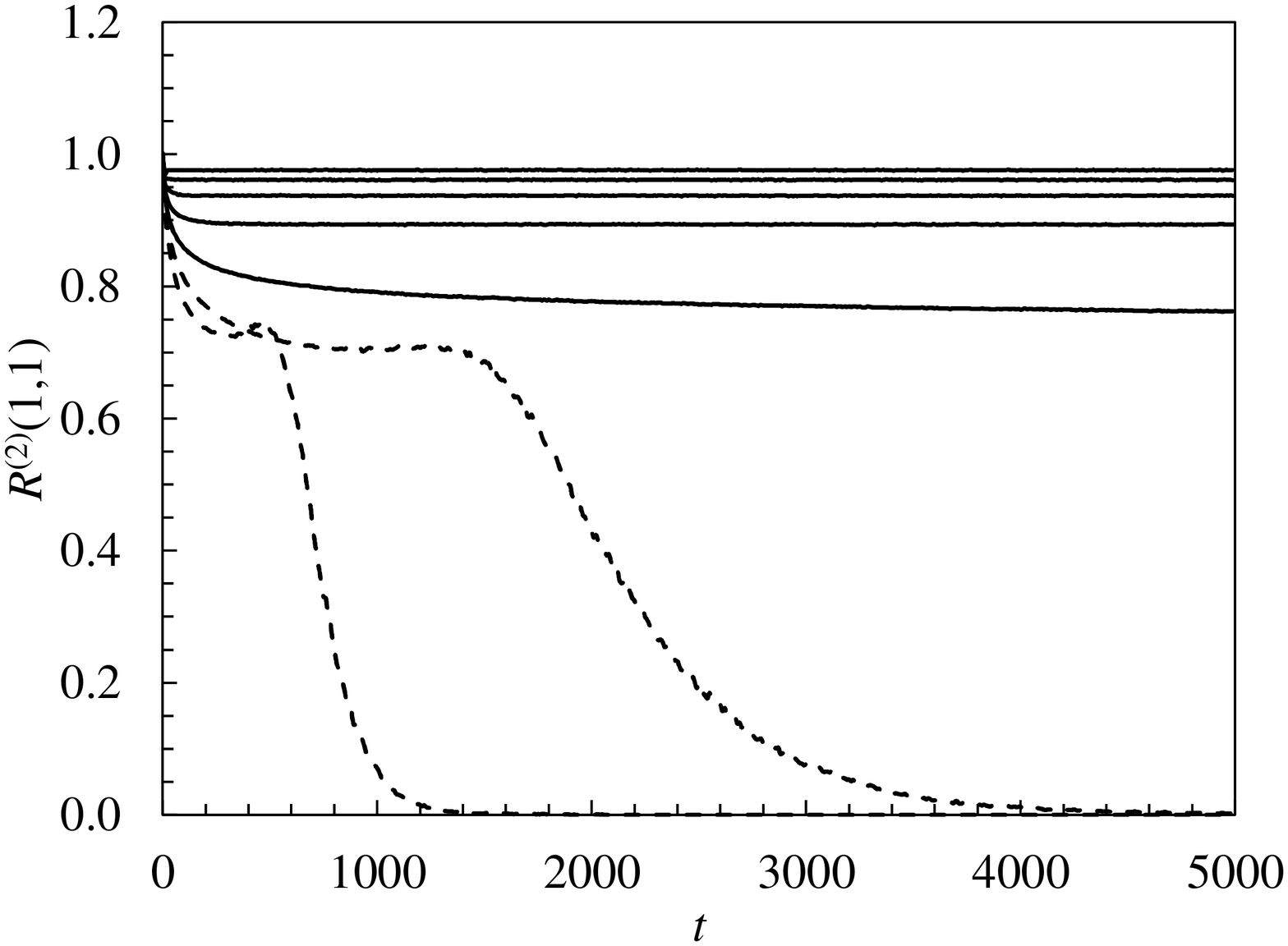} \\
\includegraphics[viewport = 60 60 780 540, scale=0.32, clip]{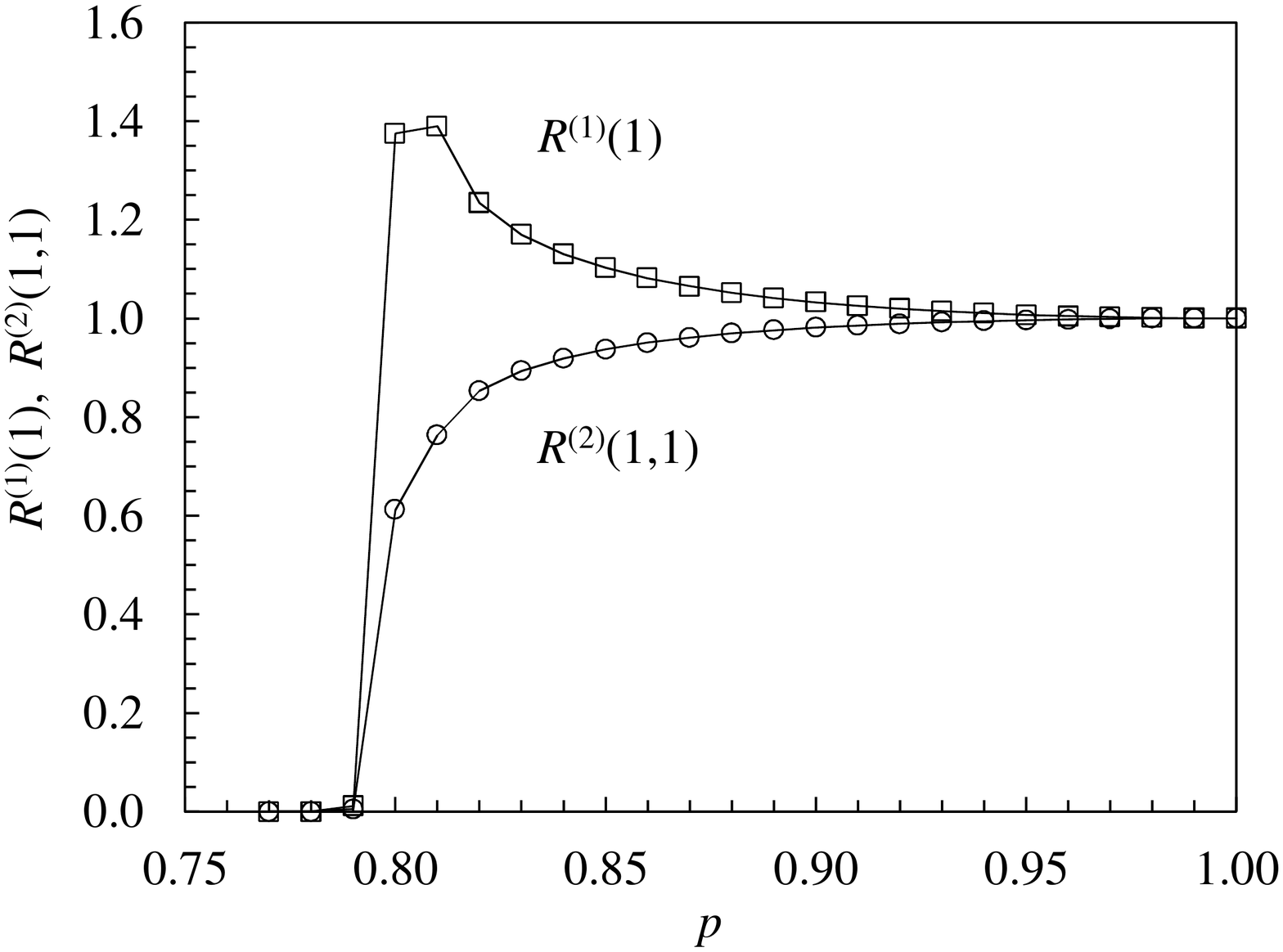} 
\end{tabular}}
\caption{\label{fig:d1d11}Ratios $R^{(1)}_{t}(1)$ (upper panel) and $R^{(2)}_{t}(1,1)$ (mid panel) for several values of $p$. Time $t$ is given in Monte Carlo steps. In the upper panel we have, from the uppermost curve down, $p=0.77$ and $0.79$ ($p<p^{*}$, dashed lines), $0.81$, $0.83$, $0.85$, $0.87$, and $0.89$ ($p>p^{*}$, solid lines). In the mid panel this order is reversed. The lower panel displays the stationary values reached by $R^{(1)}_{t}(1)$ and $R^{(2)}_{t}(1,1)$ for different values of $p$.}
\end{figure}

In this section we take an empirical look at the ratio between the $k$-cell mean field approximation $\widetilde{P}^{(k)}_{t}(x_{1}, \ldots, x_{n})$ and the exact marginal probability distribution $P_{t}(x_{1}, \ldots, x_{n})$,
\begin{equation}
\label{delta}
R^{(k)}_{t}(x_{1}, \ldots, x_{n}) = \frac{\widetilde{P}^{(k)}_{t+1}(x_{1},\ldots,x_{n})}{P_{t+1}(x_{1}, \ldots, x_{n})},
\end{equation}
as a proxy to the ``quality'' of the mean field approximation. The index $t+1$ on the right-hand side of (\ref{delta}) is a matter of convenience; we can look at $R^{(k)}_{t}(x_{1}, \ldots, x_{n})$ as a predictor for the discrepancy of the probabilities being measured in the next step of the dynamics of the PCA. Clearly, $R^{(k)}_{t}(x_{1}, \ldots, x_{n})=1$ for $k>n$, since in this case $\widetilde{P}^{(k)}_{t}(x_{1}, \ldots, x_{n})=$ $P_{t}(x_{1}, \ldots, x_{n})$. Note that $R^{(k)}_{t}(x_{1}, \ldots, x_{n})$ is related with the $n$-point equal-time connected correlation function \cite{amit} $G^{(c)}_{t}(x_{1}, \ldots, x_{n})=$ $[P_{t}(x_{1})-P(x_{1})] \cdots$ $[P_{t}(x_{n})-P(x_{n})]$, but we will not explore this relation here.

From equations (\ref{onexact}), (\ref{twoexact}) and (\ref{twoapp}) we obtain
\begin{equation}
\label{del-1}
R^{(1)}_{t}(1) = \frac{P_{t}(0)P_{t}(1)}{P_{t}(0,1)}
\end{equation}
for the discrepancy of the probability density $P_{t}(1)$ in the single-cell approximation and
\begin{equation}
\label{del-2}
R^{(2)}_{t}(1,1) = \frac{P_{t}(0,1)P_{t}(1,0)}{P_{t}(0)P_{t}(1)[P_{t}(0,1,0)+P_{t}(1,0,1)]}
\end{equation}
for the discrepancy of the probability density $P_{t}(1,1)$ in the two-cell approximation. Note that $R^{(1)}_{t}(1)$ and $R^{(2)}_{t}(1,1)$ are defined only for $p \geq p^{*}$, because for $p < p^{*}$ we have $P(1)=0$, and so every other marginal $P(\ldots, 1, \ldots)=$ $0$ as well. We define $R^{(1)}(1)=0$ and $R^{(2)}(1,1)=0$ for $p \leq p^{*}$.

To compute (\ref{del-1}) and (\ref{del-2}), we measured the probabilities $P_{t}(1)=$ $1-P_{t}(0)$, $P_{t}(0,1)=$ $P_{t}(1,0)$, $P_{t}(0,1,0)$, and $P_{t}(1,0,1)$ by Monte Carlo simulations. For each value of $p$, quantities were averaged over $2500$ realizations of the dynamics. We did not find significant differences in the data for different sizes of the array ($L=2000$, $4000$, and $8000$ cells), so we report results for $L=4000$ only. Our results appear in figure~\ref{fig:d1d11}.

At $t=0$, $R^{(1)}_{0}(1)=1$ and $R^{(2)}_{0}(1,1)=1$, because the initial state $\bm{x}^{0}$ is an uncorrelated random state with $P_{0}(x_{1}, \ldots, x_{n}) = P_{0}(x_{1}) \cdots P_{0}(x_{n})$. Otherwise, we observe rapid convergence of $R^{(1)}_{t}(1)$ and $R^{(2)}_{t}(1,1)$ to their stationary values when $p$ is away from $p^{*}$; close to $p^{*}$ correlation lengths increase and it takes longer for these quantities to reach their stationary values. We observed the empirical bounds $R^{(1)}_{t}(1) \lesssim 1.4$ and $R^{(2)}_{t}(1,1) \leq 1$ for all $p$; if we consider only the region $p>p^{*}$, then the empirical bounds read $1 \leq R^{(1)}_{t}(1) \lesssim 1.4$ and $0.6 \lesssim R^{(2)}_{t}(1,1) \leq 1$. It is thus clear that $\widetilde{P}^{(1)}_{t}(1)$ is an upper bound to $P_{t}(1)$, while $\widetilde{P}^{(2)}_{t}(1,1)$ is a lower bound to $P_{t}(1,1)$.

Our conclusion is that away from the critical point (but not necessarily very far away), the mean field approximations provide a reasonably good description of the dynamics of the PCA. In particular, even the relatively crude one-cell approximation for the density of active cells is never far off the value of the empirical density by more than $\sim 40\%$, while the two-cell approximation falls even closer to the empirical value of its respective marginal probability density.  For example, the determination of the critical point from the two-cell approximation differs from the empirical value by $\sim (2/3)/0.81 \sim 5/6$, i.e., by just $\sim 1/6$ of the empirical value. The short range of the interaction in the $p$-XOR PCA (and in PCA in general) may be partially responsible for this good agreement already at the level of the two-cell approximation.  One cannot, however, rely on the mean field approximations to determine critical exponents (they are always classical'') or phase diagrams with more than one parameter, in which case the {\it loci\/} of the critical manifolds and multicritical points may differ significantly from the ones provided by the mean field approximations \cite{marro,haye,rechtman,amit}.


\section{\label{summary}Summary and conclusions}

We estimated the critical point of the $p$-XOR PCA at $p^{*}=0.80939(3)$ and found that the model belongs to the $(1+1)$-dimensional directed percolation universality class of critical behavior. This value of $p^{*}$ agrees, within the error bounds, with some (but not all) values found before for $p_{1}^{*}$ in the DK PCA on its line $p_{2}=0$ \cite{domany,kinzel,martins,penna,weitz,kemper,rechtman}, but improves former estimates by at least one order of magintude---the best value currently available for this quantity was $p_{1}^{*}=0.8087(5)$ \cite{weitz}. Our numerical analyses took approximately 1630 CPU-hours (development and tests discounted) on Power~755 RISC processors running C/GCC code at 3.4~GHz over IBM AIX.

Our identification of the $p$-XOR PCA as an instance of the DK PCA helps to make information existent about the later available in the study of the former. It is known for some time that the diluted CA~90 (i.e., the $p$-XOR PCA) and the DK PCA are equivalent on the line $p_{2}=0$ of the later \cite{bagnoli}, but we have further identified the equivalence of the $p$-XOR with the diluted CA~102, which should thus have the same critical behavior as the DK PCA over the line $p_{2}=0$ and the diluted CA~90. The relationship between the DK PCA and two-dimensional classical spin models and the detailed knowledge of its phase diagram, that displays an extended chaotic phase, may help to classify the behavior of the $p$-XOR PCA and its siblings according to Wolfram's classes 1--4 \cite{program} (see, however, a critique of this program in \cite{gray}).

The analysis of the mean field approximations showed that the cluster approximation provided by the general prescription (\ref{mfield}) is acceptable even for clusters formed by as few as two or three cells. Clearly, the approximation is better the farther one is away from the critical point, when correlation lengths diverge. A different approach than that presented in section~\ref{assess} would be to compare the analytical expressions (\ref{onecell}) and (\ref{twostat}), and possibly higher order approximations, with their empirical counterparts. The case of $P(1)$ can be apprehended directly from figure~\ref{fig:rho}. Moreover, the ratio $R^{(1)}(1)=$ $P(0)P(1)/P(0,1)$ could perhaps be used to locate the critical point more efficiently (less computation, more precision) than with $P(1)$ alone. We intend to pursue these lines of inquire further elsewhere.

Finally, we believe that a more detailed investigation of probabilistic mixtures of deterministic CA with known properties \cite{hans,bhatta,p182q200} is worth carrying out further.


\section*{Acknowledgments}

The author thanks the S\~{a}o Paulo State Research Foundation -- FAPESP for CPU time in the LaSCADo computing cluster at FT/Unicamp through grant 2010/50646-6.


\end{document}